\documentclass[journal]{IEEEtran}
\ifCLASSOPTIONcompsoc
  \usepackage[nocompress]{cite}
\else
  \usepackage{cite}
\fi
\ifCLASSINFOpdf
\else
\fi
\usepackage{url}
\usepackage{graphicx}

\begin{document}
%
% paper title
% Titles are generally capitalized except for words such as a, an, and, as,
% at, but, by, for, in, nor, of, on, or, the, to and up, which are usually
% not capitalized unless they are the first or last word of the title.
% Linebreaks \\ can be used within to get better formatting as desired.
% Do not put math or special symbols in the title.
\title{DGEM: A New Dual-modal Graph Embedding Method in Recommendation System}
%
%
% author names and IEEE memberships
% note positions of commas and nonbreaking spaces ( ~ ) LaTeX will not break
% a structure at a ~ so this keeps an author's name from being broken across
% two lines.
% use \thanks{} to gain access to the first footnote area
% a separate \thanks must be used for each paragraph as LaTeX2e's \thanks
% was not built to handle multiple paragraphs
%
%
%\IEEEcompsocitemizethanks is a special \thanks that produces the bulleted
% lists the Computer Society journals use for "first footnote" author
% affiliations. Use \IEEEcompsocthanksitem which works much like \item
% for each affiliation group. When not in compsoc mode,
% \IEEEcompsocitemizethanks becomes like \thanks and
% \IEEEcompsocthanksitem becomes a line break with idention. This
% facilitates dual compilation, although admittedly the differences in the
% desired content of \author between the different types of papers makes a
% one-size-fits-all approach a daunting prospect. For instance, compsoc 
% journal papers have the author affiliations above the "Manuscript
% received ..."  text while in non-compsoc journals this is reversed. Sigh.

\author{Huimin~Zhou,
        Qing~Li, ~\IEEEmembership{Member,~IEEE,}
        Yong~Jiang~\IEEEmembership{Member,~IEEE,}% <-this % stops a space
        Rongwei~Yang, ~\IEEEmembership{Member,~IEEE,}
        and~Zhuyun~Qi~\IEEEmembership{Member,~IEEE}% <-this % stops a space
\thanks{\IEEEcompsocthanksitem Huimin Zhou, Graduate School at Shenzhen, Tsinghua University, Shenzhen 518055, P.R.China. Email: zhm17@mails.tsinghua.edu.cn}
\thanks{Qing Li, Peng Cheng Laboratory, Shenzhen 518055, P.R.China. Email: liq@pcl.ac.cn}
\thanks{Yong Jiang, Graduate School at Shenzhen, Tsinghua University, Shenzhen 518055, P.R.China. Email:jiangy@sz.tsinghua.edu.cn}
\thanks{Rongwei Yang, Peng Cheng Laboratory, Shenzhen 518055, P.R.China. Email: yangrw@pcl.ac.cn}
\thanks{Zhuyun Qi, Peng Cheng Laboratory, Shenzhen 518055, P.R.China. Email: qizy@pcl.ac.cn}
% <-this % stops an unwanted space
\thanks{Qing Li is the corresponding author.}
}

% note the % following the last \IEEEmembership and also \thanks - 
% these prevent an unwanted space from occurring between the last author name
% and the end of the author line. i.e., if you had this:
% 
% \author{....lastname \thanks{...} \thanks{...} }
%                     ^------------^------------^----Do not want these spaces!
%
% a space would be appended to the last name and could cause every name on that
% line to be shifted left slightly. This is one of those "LaTeX things". For
% instance, "\textbf{A} \textbf{B}" will typeset as "A B" not "AB". To get
% "AB" then you have to do: "\textbf{A}\textbf{B}"
% \thanks is no different in this regard, so shield the last } of each \thanks
% that ends a line with a % and do not let a space in before the next \thanks.
% Spaces after \IEEEmembership other than the last one are OK (and needed) as
% you are supposed to have spaces between the names. For what it is worth,
% this is a minor point as most people would not even notice if the said evil
% space somehow managed to creep in.

% The paper headers
\markboth{}%
{Shell \MakeLowercase{\textit{et al.}}: Bare Demo of IEEEtran.cls for Computer Society Journals}
\maketitle

\begin{abstract}
In the current deep learning based recommendation system, the embedding method is generally employed to complete the conversion from the high-dimensional sparse feature vector to the low-dimensional dense feature vector. However, as the dimension of the input vector of the embedding layer is too large, the addition of the embedding layer significantly slows down the convergence speed of the entire neural network, which is not acceptable in real-world scenarios. In addition, as the interaction between users and items increases and the relationship between items becomes more complicated, the embedding method proposed for sequence data is no longer suitable for graphic data in the current real environment. Therefore, in this paper, we propose the Dual-modal Graph Embedding Method (DGEM) to solve these problems. DGEM includes two modes, static and dynamic. We first construct the item graph to extract the graph structure and use random walk of unequal probability to capture the high-order proximity between the items. Then we generate the graph embedding vector through the Skip-Gram model, and finally feed the downstream deep neural network for the recommendation task. The experimental results show that DGEM can mine the high-order proximity between items and enhance the expression ability of the recommendation model. Meanwhile it also improves the recommendation performance by utilizing the time dependent relationship between items.
\end{abstract}

% Note that keywords are not normally used for peerreview papers.

% To allow for easy dual compilation without having to reenter the
% abstract/keywords data, the \IEEEtitleabstractindextext text will
% not be used in maketitle, but will appear (i.e., to be "transported")
% here as \IEEEdisplaynontitleabstractindextext when the compsoc 
% or transmag modes are not selected <OR> if conference mode is selected 
% - because all conference papers position the abstract like regular
% papers do.
%\IEEEdisplaynontitleabstractindextext
% \IEEEdisplaynontitleabstractindextext has no effect when using
% compsoc or transmag under a non-conference mode.

% For peer review papers, you can put extra information on the cover
% page as needed:
% \ifCLASSOPTIONpeerreview
% \begin{center} \bfseries EDICS Category: 3-BBND \end{center}
% \fi
%
% For peerreview papers, this IEEEtran command inserts a page break and
% creates the second title. It will be ignored for other modes.
\IEEEpeerreviewmaketitle

%\IEEEraisesectionheading{\section{Introduction}\label{sec:introduction}}

\section{Introduction}
\label{sec:introduction}
%Embedding refers to using a low-dimensional vector to represent an object. Embedding vectors have the property that objects corresponding to similar vectors in embedding space have similar meanings. 
When dealing with sparse vectors, the recommendation model based on deep learning has the problem of huge parameter magnitude for the model. For engineering, the cost of storage space is also extremely expensive. The above reasons make deep learning very inconvenient for the processing of sparse vectors, so an embedding method that can encode objects with a low-dimensional vector and still retain its properties is very important for the recommendation model based on deep learning.

In the deep learning recommendation system, the embedding method has three potential application scenarios:
\begin{itemize}
	\item As the embedding layer in the deep learning network, the embedding method can make the conversion from high-dimensional sparse feature vectors to low-dimensional dense feature vectors;
	\item As part of the preprocessing, the embedding method can generate embedding feature vectors, which are connected with other features and then used as input to the deep learning network;
	\item As one of the recall layers or recall methods of the recommendation system, it can calculate the embedding vectors similarity between users and items to generate a recommendation list.
\end{itemize}

Google published three articles of Word2Vec~\cite{Word2Vec1,Word2Vec2,Word2Vec3} for the embedding technology in 2013. Immediately afterwards, Microsoft expanded Word2Vec to Item2Vec, making it directly extend from the field of Natural Language Processing to any field that can generate sequences such as recommendations, advertisements and searching~\cite{Item2Vec}. Coincidentally, Airbnb subsequently proposed their own Item2Vec model~\cite{Airbnb}. However, the relationship between entities such as users and items are becoming more and more complex, and it is no longer a pure sequence relationship, but more a graphical data~\cite{Graph}. Therefore, applying the graph embedding methods in the recommendation system has attracted more and more attention from academic. 

DeepWalk~\cite{DeepWalk} is a preliminary model that uses random walk to convert graphical relationships into sequential relationships. Although DeepWalk can be applied to very large-scale networks, it is only suitable for unweighted graphs, but not for weighted graphs. Compared with DeepWalk's pure random walk sequence generation method, LINE~\cite{LINE} introduces the first-order and second-order proximity relationships into the objective function, which can make the final distribution of the embedding vectors of items more balanced and smoother. Node2Vec~\cite{Node2Vec} then improves DeepWalk by combining depth-first search and breadth-first search. In this way the final embedding structure can express the overall and local structure of complex graphics. Compared with Node2Vec's improvement of the walk mode, the SDNE~\cite{SDNE} model mainly solves the problem of the local structure and global structure of embedding graphics from the design of the objective function. Compared with LINE's practice of learning the local structure and the global structure separately, SDNE performs the overall optimization together, which is more conducive to obtaining the overall optimal embedding vectors. However, all the current approaches are proposed for a single environment, and they cannot be applied to real static environment and complex dynamic environment at the same time.

In this paper, we propose \textbf{\underline{D}}ual-modal \textbf{\underline{G}}raph \textbf{\underline{E}}mbedding \textbf{\underline{M}}ethod (DGEM). DGEM works in different modes according to different application scenarios; static mode (S-DGEM) in static recommendation environment and dynamic mode (D-DGEM) in dynamic recommendation environment. In this way, we solve the problem of xxx. 
In the static recommendation environment, S-DGEM extracts the graph structure model by establishing a directed weighted item graph and uses random walk of unequal probability to capture the vertex attributes of the item graph. Based on the generated item sequence data, we construct the item graph embedding vector by the Word2Vec method and feed the deep neural network for recommendation.
In the dynamic recommendation environment, D-DGEM introduces the time state to track the update of the item graph and improves the unequal probability random walk strategy in the static recommendation environment to capture the vertex attribute in the dynamic item graph. We also add auxiliary information to enhance the special solitary point expression. In this way, the timing dependence between items can be better utilized and the recommendation performance in a dynamic environment can be improved.
To improve the user experience, we further introduce the application users interests as a feature for deep learning.

To demonstrate the performance of DGEM, we construct comprehensive experiments based on the Amazon electronic product data subset. The experimental results show that 1) DGEM can use random walk to mine higher-order neighbor relationships to make up for the sparse purchasing behavior and enhance the expression ability of the model; 2) DGEM introduces the time state, which alleviates the inherent scalability and cold start problems of the recommended system.

The rest of the paper is organized as follows. In Section \ref{sec:background} we provide the analysis of background and relate work. In Section \ref{sec:frame}, we propose the framwork of DGEM. In Section \ref{sec:experiment}, analyze the performance of DGEM based on comprehensive experiments. In Section \ref{sec:conclusion}, we conclude the scheme of DGEM.

\section{Background and Related Work}
\label{sec:background}
%或者
With the continuous advancement of communication technology and network platforms, information data has shown an exponentially explosive growth trend, which has brought about serious information overload problems~\cite{IO1,IO2}. To solve this problem, the recommendation system in the applications developed by service providers has become an essential part. In real-world e-commerce applications~\cite{survey}, graph data such as social networks between users, product networks between items, and interaction networks between users and items are everywhere. Through the analysis of graph data, we can deeply understand the user's social structure, the relevance of the item, and the interaction between the user and the item. To solve the problem of graph analysis, industry and academia have proposed many methods to perform the analysis. Among them, graph embedding technology, which uses graph vertex representation method in vector space, has received more and more attention from researchers in recent years.

The graph embedding technology can be traced back to the early 2000s and its main application scenario is as a dimensionality reduction technique. The basic idea of the dimensionality reduction graph embedding techniques is to construct a set of $D$ dimensional vertices into a neighborhood-based graph at frist, and then embed the vertices into a $d$ ($d \ll D$) dimensional vector space. Laplacian Eigenmaps(LE)~\cite{LE} and Locally Linear Embedding(LLE)~\cite{LLE} are typical dimensionality reduction graph embedding techniques. The time complexity of the dimensionality reduction graph embedding algorithm is related to the square of the number of vertices, so it is suitable for small-scale graph data and has poor scalability.

Since 2010, the research direction of graph embedding technology has shifted from dimensionality reduction algorithm to scalable graph embedding algorithm. Scalable graph embedding algorithms can be divided into two categories based on factorization methods and random walk. The graph embedding algorithm based on the factorization methods(eg., Graph Factorization~\cite{GF}, GraRep~\cite{GraRep}, HOPE~\cite{HOPE}) has poor interpretability and low parallelism, and it is difficult to perform online incremental calculation. 

Random walks have been used to approximate many properties in the graph including vertex centrality~\cite{centrality} and similarity~\cite{similarity}. Among the graph embedding methods based on random walk~\cite{DeepWalk, Node2Vec, HARP, Walklets,  GenVector, DDRW, TriDNR, other}, DeepWalk is the first model to introduce deep learning. DeepWalk algorithm consists of two parts: random walk sequence generation part and parameter update part. DeepWalk can learn in parallel and can mine the local structure of the graph, but it only supports unweighted graphs, which makes it impossible to generalize. The introduction of LINE makes the type of graph no longer a limitation of the model. LINE also proposed an edge sampling algorithm to solve the limitations of classic stochastic gradient descent, which improves the sampling efficiency and effect. LINE explicitly defines a loss function to capture first-order proximity and second-order proximity, which represent first-order local relations and second-order local relations, respectively. However, it cannot be extended to the higher-order proximity. DeepWalk and LINE are essentially an algorithm based on neighborhood relations. Different sampling strategies will result in different neighborhood relations, so different vertex expressions are learned. In order to solve the problem of sampling flexibility, Microsoft proposed the Node2Vec model. Compared with rigid search methods such as DeepWalk and LINE, node2vec can control the search space by adjusting hyperparameters, thus generating a more flexible algorithm. The hyperparameter has an intuitive explanation and determines different search strategies. 

In all the proposed graph embedding methods based on random walk, they focus on the mining of vertex relationships in a static application environment. DeepWalk is a kind of pure random walk, and the vertex relationships of its mining are not biased; LINE focuses on the first-order and second-order neighbor relations; Node2Vec uses different methods to mine the different relations between vertices. Although all methods can be extended, it is essentially a static extension. In order to implement static and dynamic dual-modal scalable graph embedding method in recommendation system, we propose DGEM.

\section{Framework}
\label{sec:frame}

In this section, we introduce the design ideas of DGEM in detail, mainly including S-DGEM in static mode and D-DGEM in dynamic mode. DGEM can be roughly divided into five modules, namely item graph construction module, random walk module, solitary point processing module, graph embedding module and deep neural network module.

\subsection{Recommendation target for graph embedding}

When using the recommendation system, if the participants in the recommendation system remain relatively stable for a period of time, then we can call the recommendation system a static recommendation system (SRS). If the objects involved in the recommendation system change anytime and anywhere, then we can call the recommendation system as a dynamic recommendation system (DRS).

In a real recommendation scenario, the user's interest usually shows a diversified trend. The number of items interacted by the user accounts for a small proportion compared to the total number of items. It is difficult to train an accurate model. The traditional recommendation method has problems of insufficient interactive information mining, homogenization of recommended items and data sparseness; and in the deep learning-based recommendation method, its important embedding operations are designed for sequence data and are no longer applicable the graphic data in the real environment. In other words, the above two types of recommendation methods are not suitable for the gradually complicated and networked recommendation environment.

Therefore, in a real recommendation scenario, there are the following requirements:
\begin{itemize}
	\item The user's historical behavior record contains a large number of implicit user interest feedback behaviors such as clicks, browses, favorites, and purchases. Compared with explicit user interest feedback behaviors, the number of implicit user interest feedback behaviors is larger and more reflective of user interests preferences. Therefore, it is required that the designed recommendation method can deeply dig the user's implicit user interest feedback behavior to find the user's interest preference information.
	\item When the user uses the application, each historical behavior of the user have a time stamp, that is, the user's historical behavior has a strong order. Therefore, it is required to design the recommended method to preserve the sequence of user behavior.
	\item The recommendation system in the dynamic environment changes from moment to moment, and its change will cause the corresponding directed weighted item graph to change, which bring huge challenges to the graph embedding work. Therefore, the proposed recommendation algorithm have to seek an intermediate state for tracking and retaining changes in the item graph structure.
	\item Some items have little or no interaction with users, and it is difficult for deep neural networks to mine information. Therefore, it is required that the designed recommendation algorithm can train accurate models for these items with little interaction.
\end{itemize}

\subsection{The design of S-DGEM}

\subsubsection{The definition of related issues in S-DGEM}

\textbf{Definition 1. (Graph):} The graph consists of a finite set of non-empty vertices and a set of edges between vertices, usually expressed as $G(V, E)$, where $G$ represents a graph, $V=\{v_1, v_2, ..., v_n\}$ is a set of vertices in graph $G$, and $E = {\{e_{ij}\}}_{i, j = 1}^n$ is a set of edges in graph $G$.

If the edge $e_{ij}$ between the vertices $v_i$ and $v_j$ has no direction, the edge $e_{ij}$ is called an undirected edge, otherwise the edge $e_{ij}$ is called a directed edge. If all edges in the graph are undirected edges, we call the graph undirected graph. Similarly, if all edges in the graph are directed edges, we call the graph directed graph.

Some edges of the graph may have numbers associated with it. We generally call this kind of number related to the edge weight. These weights can represent the distance or cost from one vertex to another. Such a graph with weights is usually called a weighted graph. In the weighted graph, we record the weight of the edge $e_ {ij}$ between the vertices $v_i$ and $v_j$ as $w_{ij}$. The value of the weight $w_{ij}$ is usually non-negative. If the edge $e_ {ij}$ exists, $w_{ij}> 0$, otherwise $w_{ij} = 0$. In general, we record the weighted graph as $G (V, E, W)$.

\textbf{Definition 2.(Graph Embedding):} Given a graph $G(V, E)$, the essence of graph embedding is a mapping $f: v_i \rightarrow y_i \in R^d$, $\forall i \in \{1,2, ..., n\}$, where $d \ll |V|$, and the function $f$ retains some proximity defined on the graph $G$. That is, graph embedding maps each vertex to a low-dimensional feature vector space, and attempts to preserve the connection strength relationship between vertices.

There are two pairs of vertices $(v_i, v_j)$ and $(v_i, v_k)$. If their proximity is related to the connection strength, suppose there is a connection strength relationship $w_ {ij}> w_ {ik}$. Then in this case, after mapping into the embedding space, the distance between $v_i$ and $v_j$ is closer than the distance between $v_i$ and $v_k$.

\begin{table}[!t]
	\renewcommand{\arraystretch}{1.3}
	\caption{Notations of S-DGEM}
	\label{table1}
	\centering
	\begin{tabular}{c|c}
		\hline
		\bfseries Notation & \bfseries Description \\
		\hline
		$G$ & The graphical representation of data \\
		$V$ & The set of vertices in graph $G$\\
		$E$ & The set of edges in graph $G$\\
		$W$ & The set of weights of edges in graph $G$\\
		$n$ & The number of vertices in graph $G$\\
		$v_i$ & One of vertices in graph $G$\\
		$e_{ij}$ &  The edge between the vertices $v_i$ and $v_j$\\
		$w_{ij}$ & The weight of the $e_{ij}$\\
		$d$ & The dimension of embedded space \\
		\hline
	\end{tabular}
\end{table}

\subsubsection{The construction of directed weighted item graph}
After obtaining the standard user historical behavior sequence, we can proceed to build a directed weighted item graph. In the standard historical behavior sequence of the same user, we record two consecutive items $item_i$ and $item_j$ as the item pair $(item_i, item_j)$. For each item pair $(item_i, item_j)$, if there are no edge $e_{ij}$ between the two items $item_i$ and $item_j$, then add the edge $e_{ij}$, $e_{ij}$ is a directed edge, the direction of $e_{ij}$ is the item with the earlier timestamp to the item with the later timestamp, and the weight of $e_{ij}$ is recorded $w_{ij}$ is 1; if an edge $e_ {ij}$ already exists between the two items $item_i$ and $item_j$ in the item pair, the edge is no longer added, and the original weight $w_{ij}$ of the edge plus 1. For the edge $e_{ij}$ weight $w_{ij}$ operation of the item pair $(item_i, item_j)$, the $w_{ij}$ is organized into a mathematical form as follows:
\begin{equation}
w_{ij}=\left\{
\begin{array}{rcl}
w_{ij} + 1    &      & \exists e_{ij}\\
1       &      & otherwise\\
\end{array} \right.
\end{equation}
Specifically, the weight of the final directed weighted item graph is equal to the number of occurrences of related item pairs in the historical purchase behavior of all users, that is, the weight $w_{ij}$ is equal to The frequency of conversion of $item_i$ to $item_j$ in the purchase history of all users. The directed weighted item graph constructed in this way can preserve the context of the items in the user's purchase history and the similarities between different items.

\begin{figure}[!t]
	\centering
	\includegraphics[scale=0.2]{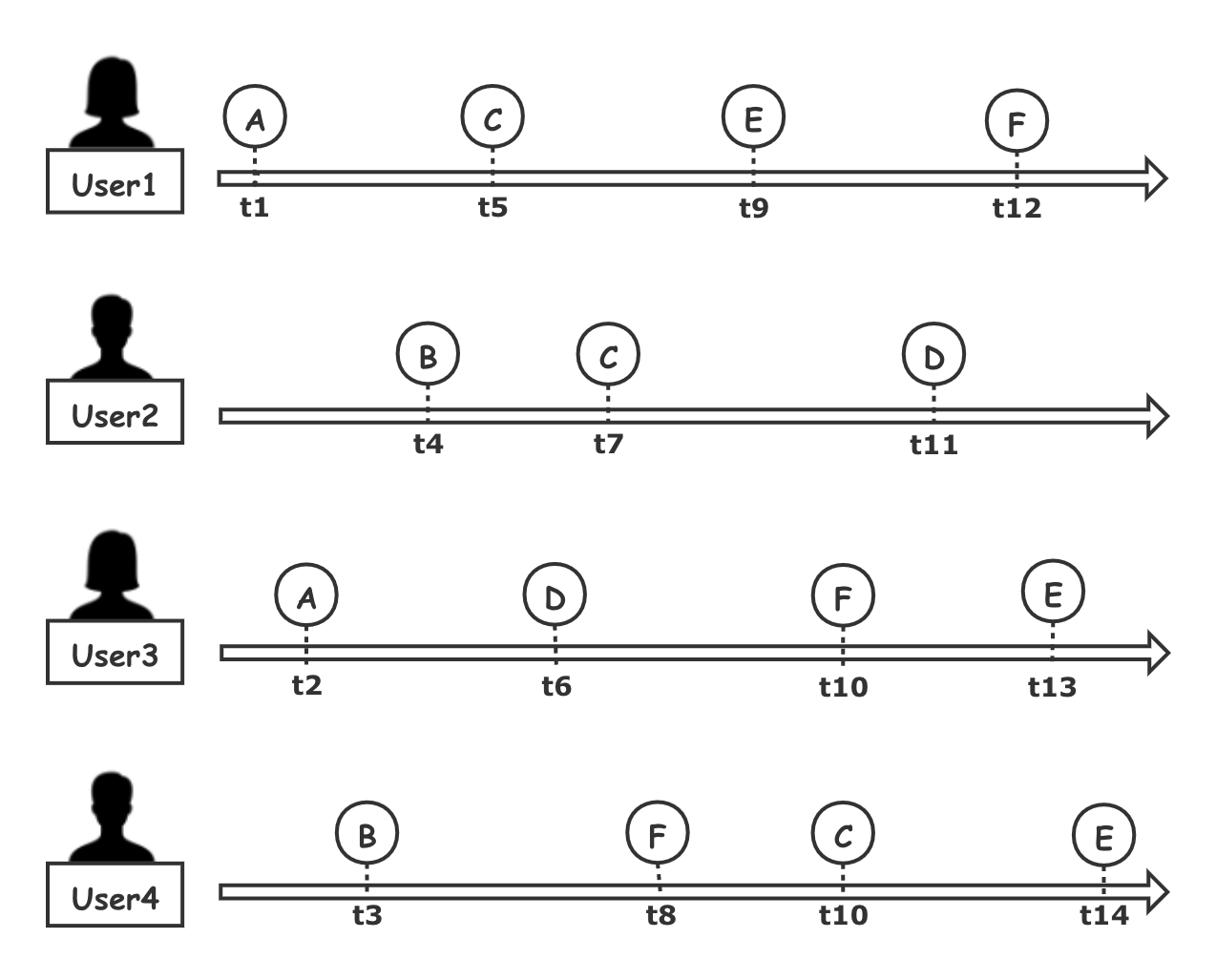}
	\caption{The historical purchase behavior sequence of 4 users, where each arrow represents a user’s complete historical purchase behavior sequence, the letters in the circle above the arrows indicate items that have interacted with the user, and the numbers below the arrows indicate the time when the user interacts with the item.}
	\label{fig1}
\end{figure}

\begin{figure}[!t] % use float package if you want it here
	\centering
	\includegraphics[scale=0.2]{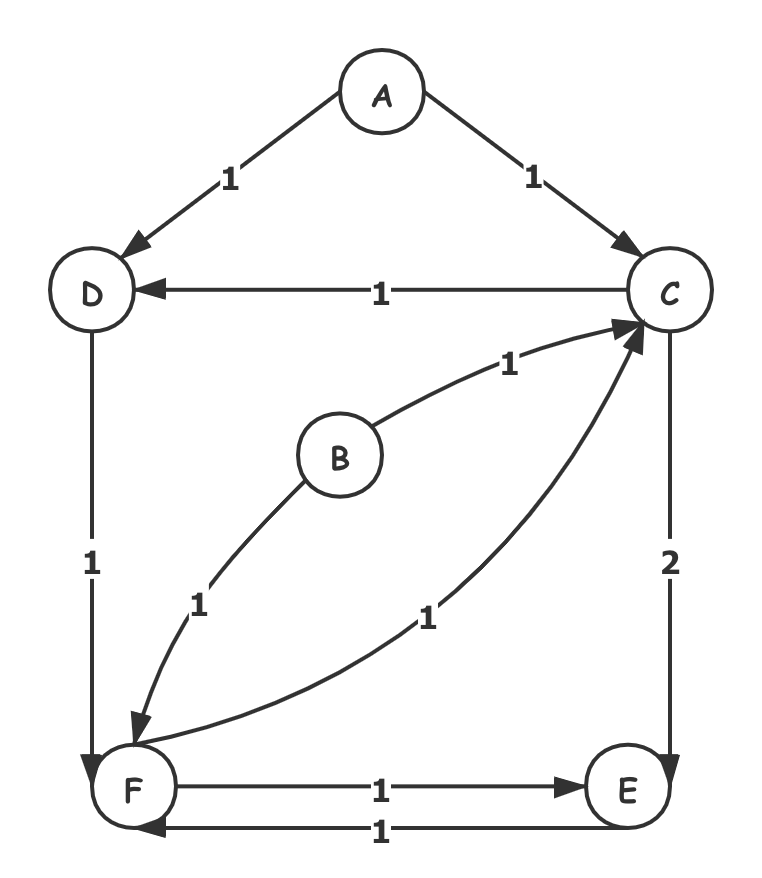}
	\caption{The directed weighted item graph constructed according to the historical purchase behavior sequence of 4 users shown in Fig.~\ref {fig1}}
	\label{fig2}
\end{figure}

Fig.~\ref {fig1} shows the historical purchase behavior sequence of 4 users. The historical purchase sequence of $user1$ is $ACEF$, the $user2$ is $BCD$, the $user3$ is $ADFE$ and the $user4$ is $BFCE$. For $user1$, her historical purchase behavior item pairs are $\{(A, C), (C, E), (E, F)\}$, because there is no edge between these item pairs, so we need to add directed edges to them and set their weights to 1. By analogy, the historical purchase behavior items of $user2$, $user3$ and $user4$ are also added with directed edges and weights are modified. Fig.~\ref {fig2} is a directed weighted item graph constructed according to the historical purchase behavior sequence of 4 users shown in Fig.~\ref {fig1}. It is worth noting that the item pair $(C, E)$ appears twice in the historical purchase behavior of all users, so the weight of directed edge $e_{CE}$ is 2. Another interesting thing is that there is a directed edge between the item pair $(E, F)$ and the item pair $(F, E)$, forming a closed loop.

\subsubsection{The design of random walk with unequal probability}
In the abstract conceptual model of random walk, it may be difficult to predict the occurrence of a single random event, but we can confirm the distribution of a large number of random events. That is to say, in the face of a single random event, we may predict that there may be a difference in what happens, but in the face of a large number of random events, we can predict the overall feature similarity. Therefore, random walk can be used to capture the topology of directed weighted item graphs. As the name implies, random walk can select a vertex in the graph as the first step, and then randomly move on the edge. Truncated random walks define the maximum length of all walk sequences.

\begin{figure}[!t] % use float package if you want it here
	\centering
	\includegraphics[scale=0.2]{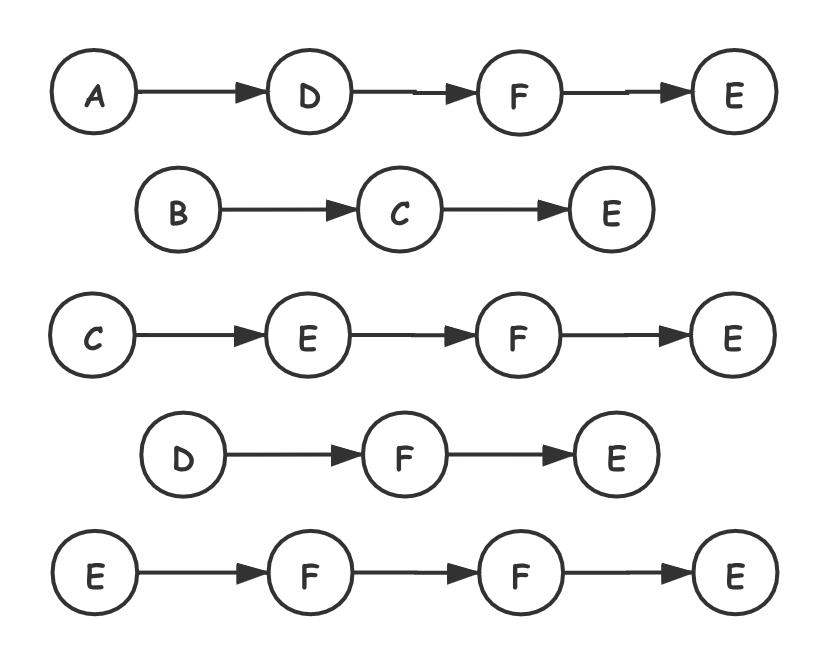}
	\caption{The item sequences generated by random walk with unequal probability according to the directed weighted item graph shown in Fig.~\ref{fig2}}
	\label{fig3}
\end{figure}

In the static recommendation environment, after constructing a directed weighted item graph, we randomly sort all vertices in the vertex set $V$ in the graph for $T$ times. For each randomly sorted vertex sequence, we use each vertex as the starting vertex of random walk according to the vertex sequence, and transfer to the adjacent vertex according to the transition probability $Pr(v_j|v_i)$ until the random walk length meets the requirements. The transition probability of random walk with unequal probability is the proportion of the weights of adjacent connected edges, that is to say, the transition probability of large edge weight will be higher. The mathematical expression of the transition probability of random walk with unequal probability is:
\begin{equation}
Pr(v_j|v_i)=\left\{
\begin{array}{rcl}
\alpha\frac{w_{ij}}{\sum_{j' \in o(v_i)}{w_{ij'}}}    &      &v_i \neq v_j\\
(1-\alpha)+\alpha\frac{w_{ij}}{\sum_{j' \in o(v_i)}{w_{ij'}}}     &      & v_i = v_j \\
0       &      & other\\
\end{array} \right.
\end{equation}
Among them, $o(v_i)$ represents the set of all directed edges that go out from the vertex $v_i$, and $\alpha$ represents the hyperparameter of whether to stay at the current vertex. Fig.~\ref{fig3} is generated by random walk with unequal probability according to the directed weighted item graph shown in Fig.~\ref{fig2}.

\subsubsection{The generation of graph embedding}
\begin{figure}[!t] % use float package if you want it here
	\centering
	\includegraphics[scale=0.2]{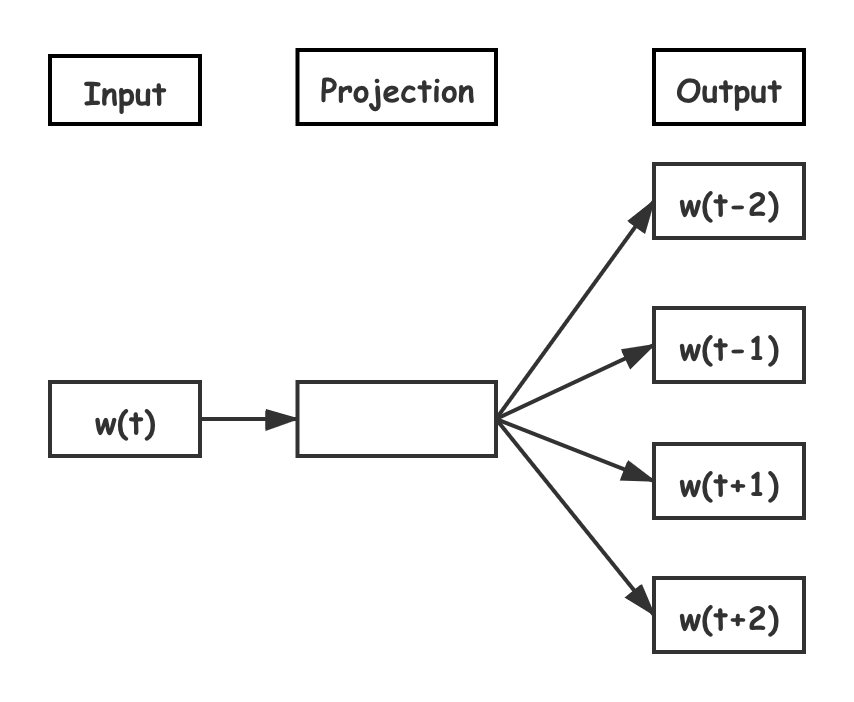}
	\caption{The model of Skip-Gram. Skip-Gram model can be expressed as a shallow neural network composed of input layer, projection layer and output layer. Skip-Gram model predicts the generation probability of each word in the context based on the current word. Among them, $w(t)$ is the word of current concern, $w(t-2)$, $w(t-1)$, $w(t+1)$,$w(t+2)$ are the words that appear in the content. Here, the size of the sliding window is set to 2.}
	\label{fig4}
\end{figure}

Word2Vec model is suitable for generating embedded vectors of serial data. Then, after we get the item sequence generated by random walk, we can naturally use the Word2Vec model to obtain the graph embedding vertices of items. Here, we use the Skip-Gram model to learn the graph embedding vertices of items, as shown in Fig.~\ref{fig4}, then its goal is to maximize the simultaneous occurrence probability of two vertices in the obtained sequence, that is, the mathematical expression of our optimization goal is:
\begin{equation}
max_f\,\log Pr(I_T=\{v_{i-w}, ..., v_{i+w}\}\backslash v_i|f(v_i))
\end{equation}
Where $w$ is the window size of the context in the sequence of items generated by random walk. If it is assumed that the vertices of the item is independent of each other, then we can get the following equation:
\begin{equation}
Pr(I_T\backslash v_i|f(v_i))\\
=\prod_{v_k\in I_T}Pr(v_k|f(v_i))
\end{equation}
Because the original Skip-Gram model iteration speed is too slow, we introduced a negative sampling method to accelerate the item's graph embedding training, then the mathematical expression of our optimization goal can be written in the following form:
\begin{equation}
min_f \log \sigma(f(v_j)^Tf(v_i)) + \sum_{t\in N(v'_i)}\log\sigma(-f(v'_t)^Tf(v_i))
\end{equation}
Among them, $N (v'_i)$ is the negative sample of $v_i$, and $\sigma (·)$ is the sigmoid function:
\begin{equation}
\sigma(x) = \frac{1}{1+e^{-x}}
\end{equation}
From experience, the larger $|N (v'_i)|$, the better the effect.

\subsection{The design of D-DGEM}

\subsubsection{The definition of related issues in DGE}

\textbf{Definition 3.(Dynamic time graph):} Given a dynamic time graph $G_T(V, E_T, \tau)$, where $V=\{v_1, v_2, ..., v_n\}$ is the set of vertices in graph $G_T$, and $E_T$ is a set of edges with time label in graph $G_T$, $\tau$ is a function that maps the time label of edge attached to a timestamp. For convenience, without special instructions, we default $\tau$ as the conversion function of UnixTime and real time directly.

\textbf{Definition 4.(Temporal walk):} Given a dynamic time graph $G_T(V, E_T, \tau)$, we record the temporal walk from vertices $v_1$ to $v_k$ as a sequence of vertices, written in the mathematical form: $<v_1, ..., v_m, ..., v_k>$, where $\forall i \in [1, k-1 ]$, $<v_i, v_{i + 1}> \in E_T$ is always established and satisfied $\tau(v_ {i-1}, v_i) \leq \tau (v_i, v_{i + 1})$ such a strict timing relationship. For two arbitrary vertices $v_i$ and $v_j$ in the vertex set $V$, if there is a temporal walk from the vertex $v_i$ to the vertex $v_j$, then we can consider the vertex $v_i$ and the vertex $v_j$ to be temporal connected.

The definition of temporal walk echoes the standard definition of random walk in the directed weighted graph of the static recommendation method, except that there is one more constraint that the walk must follow the timing relationship, that is, the time label of the passing edge must be incremental.

\textbf{Definition 5.(Dynamic graph embedding):} Given a dynamic time graph $G_T(V, E_T, \tau)$, the essence of dynamic graph embedding is a kind of mapping $f: v_i \rightarrow y_i \in R^d$, $\forall i \in \{1,2, ..., n\}$, where $d \ll | V |$, the mapped function $f$ indicates that the vertices in the graph $G_T$ are mapped to $d$ dimensional representation vectors suitable for downstream machine learning tasks.

\begin{table}[!t]
	\renewcommand{\arraystretch}{1.3}
	\caption{Notations of S-DGEM}
	\label{table2}
	\centering
	\begin{tabular}{c|c}
		\hline
		\bfseries Notation & \bfseries Description \\
		\hline
		$G_T$ & The graphical representation of data with time label\\
		$V$ & The set of vertices in graph $G_T$\\
		$E_T$ & The set of edges with time label in graph $G_T$\\
		$\tau$ & The function of mapping time label to timestamp\\
		$n$ & The number of vertices in graph $G_T$\\
		$v_i$ & One of vertices in graph $G_T$\\
		$e_{ij}$ &  The edge between the vertices $v_i$ and $v_j$\\
		$wf_{ij}$ & The frequency weight of the $e_{ij}$\\
		$wt_{ij}$ & The time weight of the $e_{ij}$\\
		$d$ & The dimension of embedded space \\
		\hline
	\end{tabular}
\end{table}

\subsubsection{The enhancement of directed weighted item graph}
After obtaining the standard user historical behavior sequence, we can proceed to build a dynamic directed weighted item graph. In the standard historical behavior sequence of the same user, we record two consecutive items $item_i$ and $item_j$ as the item pair $(item_i, item_j)$. For each item pair $(item_i, item_j)$, if there are no edge between the two items $item_i$ and $item_j$, then add the edge $e_{ij}$, $e_{ij}$ is a directed edge, the direction of $e_{ij}$ is the item with the earlier timestamp to the item with the later timestamp, and the frequency weight of $e_{ij}$ is recorded $wf_{ij}$, $wf_{ij}$ is assigned a value of 1, and the later timestamp is added to the list of time weight $wt_{ij}$. If there is already an edge $e_{ij}$ between $item_i$ and $item_j$, the new edge $e_{ij}$ no longer be added, and the original frequency weight $wf_{ij}$ of $e_{ij}$ is increased by 1 , and the later timestamp added to the list of time weights $wt_{ij}$. For the edge $e_{ij}$ weight $w_{ij}$ operation of the item pair $(item_i, item_j)$, the frequency weight $wf_{ij}$ is organized into a mathematical form as follows:
\begin{equation}
wf_{ij}=\left\{
\begin{array}{rcl}
wf_{ij} + 1    &      & \exists e_{ij}\\
1       &      & otherwise,\\
\end{array} \right.
\end{equation}
The time weights $wt_{ij}$ is organized into a mathematical form as follows:
\begin{equation}
list[wt_{ij}]=\left\{
\begin{array}{rcl}
list[wt_{ij}] + wt_{ij}     &      & \exists e_{ij}\\
wt_{ij}       &      & otherwise,\\
\end{array} \right.
\end{equation}
Specifically, the frequency weight $wf_{ij}$ of the directed weighted item graph is equal to the number of occurrences of related item pairs in the historical purchase behavior of all users, that is, the frequency weight $wf_{ij}$ of edge $e_{ij}$, it is equal to the frequency of conversion of $item_i$ to $item_j$ in the purchase history behavior of all users, and the time weight $wt_{ij}$ is the timestamp set of $item_i$ converted to $item_j$ in all user purchase history behaviors.

\begin{figure}[!t] % use float package if you want it here
	\centering
	\includegraphics[scale=0.2]{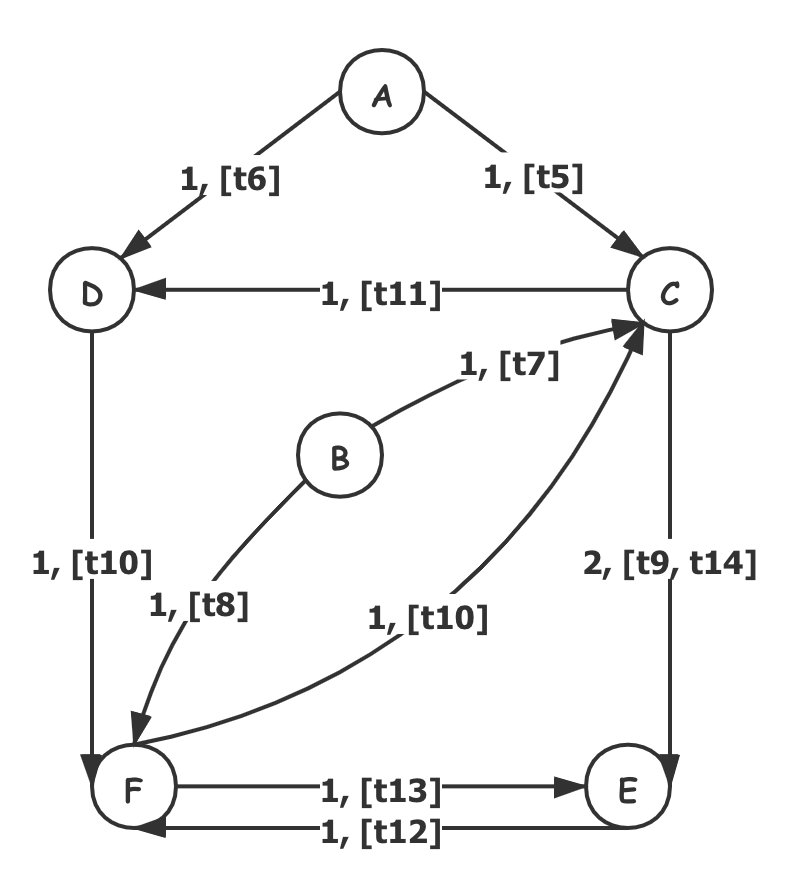}
	\caption{The dynamic directed weighted item graph constructed according to the historical purchase behavior sequence of 4 users shown in Fig.~\ref {fig1}.}
	\label{fig5}
\end{figure}

Fig.~\ref{fig1} shows the historical purchase behavior sequence of 4 users, the historical purchase sequence of $user1$  is $ACEF$, the $user2$ is $BCD$, the $user3$ is $ADFE$ and the $user4$ is $BFCE$. For $user1$, her historical purchase behavior item pairs are $\{(A, C), (C, E), (E, F)\}$, because there is no edge between these item pairs, we need to add directed edges to them, set their frequency weight to 1, and add the later timestamp to the time weight list of the edges. By analogy, the historical purchase behavior items of $user2$, $user3$ and $user4$ are also added with directed edges, frequency weights are modified and time weights are added. Fig.~\ref {fig5} is a dynamic directed weighted item graph constructed according to the historical purchase behavior sequence of 4 users shown in Fig.~\ref {fig1}. It is worth noting that the item pair $(C, E)$ appears twice in the historical purchase behavior of all users, so the frequency weight of the directed edge $e_ {CE}$ is 2, and the time weight list contains two time stamps. Another interesting point is that there is a directional edge between the item pair $(E, F)$ and the item pair $(F, E)$, forming a closed loop.

\subsubsection{The design of random walk strategy based on tense}

In the static recommendation environment, random walk is an algorithm that ignores time labels. The main purpose of the algorithm is to perform a fixed-length random walk on each vertex in the vertex set $V$ in order to collect a sufficient number of item sequences. In the dynamic recommendation environment, for the dynamic time graph $G_T(V, E_T, \tau)$, the temporal walk from vertex $v_1$ to $v_k$ is a form of vertex sequence, that is, $<v_1, ..., v_m, ..., v_k>$, where $\forall i \in [1, k-1], e_{i(i+1)} \in E_T$ is always established and satisfies $\tau (v_{i-1}, v_i) \leq \tau (v_i, v_{i + 1})$ such a strict timing relationship is always established, which shows that the temporal walk requires not only a start vertex $v_{start}$, but also a start time $t_{start}$.

In the dynamic time graph, each edge $e_ {ij} = (v_i, v_j) \in E_T$ is related to time $t = \tau(e_ {ij}) = \tau(v_j)$. The selection of the starting vertex can also be considered as the selection of the starting edge. We can find a timestamp according to the uniform distribution or the weighted distribution, and then find the edge closest to the timestamp as our starting edge. There are two types of edge selection: unbiased and biased. The selection of the start edge is expressed in mathematical form as:
\begin{equation}
Pr(e_{start})=\left\{
\begin{array}{rcl}
\frac{1}{|E_T|}     &      & Unbaised \\
\frac{\exp[\tau(e_{start})-t_{min}]}{\sum_{e'\in{E_T}} \exp[\tau(e')-t_{min}]}       &      & Baised\\
\frac{\eta(e)}{\sum_{e'\in{E_T}}\eta(e')}       &      & Baised\\
\end{array} \right.
\end{equation}
Uniform distribution is an unbiased start edge selection strategy. Its essence is to select an edge from the edge set $E_T$ with equal probability. Both exponential and linear distributions are biased starting edge selection strategy.

The selection of the start edge of the temporal walk is a very advantageous method, because this is a method of time-biasing the temporal walk. Therefore, when performing the time series regression task or time series classification task downstream, the time-biased method for temporal walk can improve the prediction performance. Not only is the selection of the start edge unbiased or biased, the transition probability of the edge can also be divided into two categories when performing temporal walks. The transition probability of edges is written in the mathematical form as:
\begin{equation}
Pr(e_{next})=\left\{
\begin{array}{rcl}
\frac{1}{|NE_T|}     &      & Unbaised, \\
\frac{\exp[\tau(e_{next})-\tau{(e_{cur})}]}{\sum_{e'\in{NE_T}} \exp[\tau(e')-\tau{(e_{cur})}]}       &      & Baised\\
\frac{\eta(e)}{\sum_{e'\in{NE_T}}\eta(e')}       &      & Baised\\
\end{array} \right.
\end{equation}
Among them, $NE_T$ is expressed as a set of all directed edges out of one vertex $v_j$ of edge $e_ {cur} = (v_i, v_j)$. Uniform distribution is an unbiased selection strategy of adjacent edges. Its essence is to select an edge with equal probability from the set of adjacent edges. Both exponential and linear distributions are biased adjacent edge selection strategy. If the function $\tau(·)$ in the exponential distribution is a monotonically increasing function, then the exponential distribution is a strategy for selection of adjacent edges that tends to appear later; if the function $\tau(·)$ is a monotonically decreasing function, then exponential distribution is an adjacent edge selection strategy that favors the selection of successive occurrence edges. $\eta(·)$ in a linear distribution is a function, which itself is a strategy for selection of adjacent edges that favors the selection of successive occurrence edges.

Given a dynamic time graph $G_T (V, E_T, \tau)$, let $S$ be the set of all possible random walks on $G_T$, and let $S_T$ be the set of all possible temporal walks on $G_T$. It is easy to see that the temporal walk set $S_T$ is included in the random walk set $S$, and $S_T$ occupies only a small part of $S$. The temporal walk we proposed can be regarded as a random walk that samples a set of random walks strictly following the timing relationship from the random walk set $S$. In some cases, the random walk order that we sample may be invalid if it does not observe time dependence. For example, suppose that each edge represents an interaction event between two people (for example, the purchase behavior after sharing a shopping link), then temporal walk can be regarded as an effective time-based path.

\subsubsection{The processing of solitary point}

The cold start item, which is a item without user interaction, is represented in a directed weighted graph as a solitary point. Learning accurate embedding vector for cold start items is still a challenge. To solve the cold start problem, we use auxiliary information attached to the cold start item to enhance the item's graph embedding. In general, items with similar auxiliary information should be closer in the embedding space. Based on this assumption, we propose a method for embedding auxiliary information. We use $H$ to denote the embedding matrix of items or auxiliary information. Specifically, $H_i^0$ represents an embedding vector of ${item}_i$, and $H_i^c$ represents embedding $cth$ type auxiliary information attached to ${item}_i$. Then, for items with $n$ kinds of auxiliary information, we have $n+1$ vectors $H_i^0, H_i^1, ..., H_i^n \in R^d$, where $d$ is the embedding dimension. Based on experience, we set the dimension of the embedding vector of the item and the embedding vector of the auxiliary information to be the same. To merge auxiliary information, we connect the $n + 1$ embedding vectors of ${item} _i$ and add a layer with an average pool operation to summarize all embeddings related to ${item} _i$:
\begin{equation}
S_i = \frac{1}{n+1}\sum_{c=0}^nH_i^c
\end{equation}
Where $S_i$ is the aggregate embedding of ${item} _i$. In this way, we merge the auxiliary information so that items with similar auxiliary information are closer in the embedding space.

\subsection{The architecture of deep neural network}

Static mode and dynamic mode share the same deep neural network module. The recommended model we adopt is a basic framework based on embedded and multi-layer perceptrons. 

The graph embedding module we designed is used as a part of preprocessing. The function is to pre-train the embedding feature vector of the item, and connect it with other feature vectors before using it as the input of the deep learning network. After obtaining the dense overall representation vector, we use the fully connected layer to automatically learn the combined features. 

\begin{figure}[!t]
	\centering
	\includegraphics[scale=0.18]{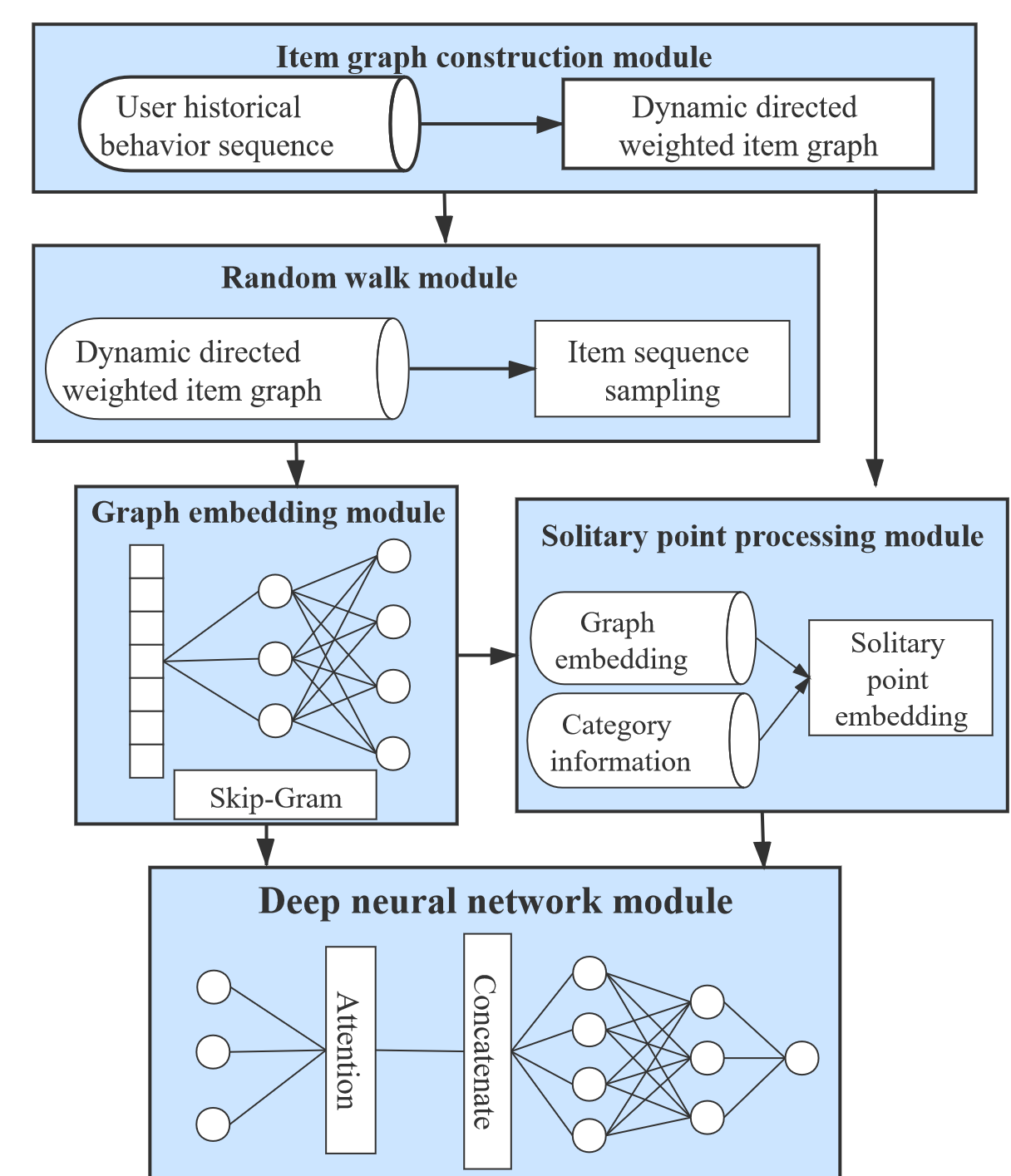}
	\caption{The implementation diagram of dual-modal deep learning recommendation scheme system based on graph embedding mechanism.}
	\label{fig6}
\end{figure}

Among them, we add an attention mechanism. When processing, the deep neural network with attention mechanism value the relevant behavior history, and the irrelevant history can even be ignored. It is also intuitive to reflect such ideas into the model. If we follow the previous approach, we consider that all behavior records have the same effect. Corresponding to the model, we use an average pooling layer to average the embedding vectors of all products interacted by the user to form the user vector of this user. Or add a timestamp to make the impact of the latest behavior greater, corresponding to the model is to adjust the weight according to time when doing average pooling. In the traditional attention mechanism, given two item embedding vectors, such as $u$ and $v$, usually do the dot product $uv$ or $uWv$ directly, where $W$ is a weight matrix of $| u | \times | v |$. But we have made further improvements, focusing on the attention mechanism shown in the Fig.~\ref{fig6}. First of all, the corresponding element difference vectors of $u$, $v$, and $u-v$ are combined as input, and then fed to the fully connected layer, and finally obtained Weights. 

In the training process, the objective function we use is a negative log-likelihood function, which is defined as:
\begin{equation}
Loss = -\frac{1}{N}\sum_{(x,y)\in Train}[y\log p(x)+(1-y)\log(1-p(x))]
\end{equation}
Among them, $Train$ is the training set of size $N$, $x$ is the input of the deep neural network, $y \in \{0,1\}$ is the attribute label, and $p(x)$ is the output of the deep neural network after the softmax layer. The trained deep neural network will be used to predict the user's next item of interest.

\section{Experiment}
\label{sec:experiment}

\subsection{Data Set}
In order to verify the performance of DGEM, we adopted the Amazon public dataset as the benchmark dataset. The data set contains a product review data set and a product information data set. The product review data set includes a total of 142.8 million reviews on the Amazon website from May 1996 to July 2014. We finally selected a subset of electronic product data from Amazon’s public data set for experiments. The Amazon Electronics product data subset also contains the product review data set (reviews$\_$Electronics) and product information data set (meta$\_$Electronics).The product review data set contains information such as reviewer ID, product ID, review usefulness rating, review text, review time, etc. The product review data set contains product ID, product name, Product pictures, product category lists and product descriptions.

In the static recommendation environment, for the product review data set in the Amazon electronic product data subset, we only retain the three types of information: reviewer ID (reviewerID), product ID (asin), unix review time (unixReviewTime). As for the product information in the Amazon electronic product data subset,  we only keep the two types of information: product ID (asin) and product category (categories). Considering that some product IDs appear in the product information data set but not appear in the product review data set, we have to preprocess the product information data set, that is, only the product ID that appeared in the product review data set is retained in the product information data set and the duplicates are removed. After preprocessing, the Amazon electronic product data subset contains only 192403 users, 63001 products, 801 categories and 1689188 samples. Each reviewer has published at least 5 reviews, and each product has at least 5 reviews.

In the dynamic recommendation environment, different from the data set preprocessing in static environment, considering that some product IDs appear in products in the product information data set but not in the product review data set, we need to preprocess the product information data set, that is, product IDs which appear in the product information data set but not in the product review data set are extracted separately as soliraty points, and product IDs that have appeared in both product information data set and  product review data, the duplicates are removed. We also remove nearly a third of the edge connections between items with later time stamp to simulate a dynamic recommendation environment.

\subsection{Comparison method}
We compare DGEM with the following algorithms:
\begin{itemize}
	\item LR~\cite{LR}. Logistic Regression (LR) is an online algorithm for generalized linear models and the basic model before the rise of deep learning.
	\item BaseModel. BaseModel follows the Embedding-MLP architecture and is the basis for most of the deep network development for subsequent click-through modeling. It has laid a solid foundation for our model comparison.
	\item Wide\&Deep~\cite{WideDeep}. The Wide\&Deep model is a model for joint training of shallow and deep models. Among them, the shallow model is a basic linear model that is mainly used to obtain cross-features. The deep model is actually a feed-forward neural network that is mainly used for feature generalization.
	\item PNN~\cite{PNN}. The PNN model is designed with a Product Layer, which realizes the purpose of extracting linear features and nonlinear features together. The linear features are obtained through the embedding layer, while the nonlinear features are obtained through the inner product and the outer product.
	\item DeepFM~\cite{DeepFM}. In the DeepFM model, the factorization machine part is responsible for feature extraction of the first-order features and the second-order features formed by combining the first-order features in pairs, and the multi-layer perceptron is responsible for fully connecting the input first-order features. The formed high-order features are used for feature extraction. In other words, the DeepFM model combines the advantages of the depth model and the breadth model. By factoring and multi-layer perceptron sharing feature embedding vectors for joint training, the purpose of learning both low-order feature combinations and high-order feature combinations can be achieved. This end-to-end model eases the pressure of feature engineering.
	\item DeepWalk~\cite{DeepWalk}. Use random walk to transform networked relationships into sequential relationships for subsequent processing.
	\item LINE~\cite{LINE}. By introducing the first-order and second-order proximity relations into the objective function, the distribution of embedding items finally learned can be more balanced and smoother.
	\item Node2Vec~\cite{Node2Vec}. DeepWalk has been improved, combining depth-first search and breadth-first search, so that the final embedded structure can express the overall and local structure of the network.	
\end{itemize}

\subsection{Evaluation index}
In the field of binary classification, AUC is a widely used indicator, and its calculation formula is:
\begin{equation}
AUC = \frac{\sum_{i\in positiveClass}Rank_i-\frac{N_{positive}(1+N_{positive})}{2}}{N_{positive}N_{negative}}
\end{equation}
In order to better use the AUC index to measure our model, we introduced the user weighted AUC variation, which judges the model's merits by averaging the AUC of each user. The mathematical expression of user weighted AUC is:
\begin{equation}
GAUC = \frac{\sum_{i=1}^n impression_i\times AUC_i}{\sum_{i=1}^n impression_i}
\end{equation}
In order to better evaluate the relative improvement between various models, we also introduced the RelaImpr indicator, whose mathematical expression is:
\begin{equation}
RelaImpr = \big(\frac{AUC(measured model)-0.5}{AUC(base model)-0.5}-1\big)\times 100\%
\end{equation}

\subsection{Experimental setup}
The main part of DGEM is deployed on the e-commerce website, so that both the static recommendation environment and the dynamic recommendation environment can be obtained at the same time. The realization of the whole scheme consists of item graph construction module, random walk module, solitary point processing module, graph embedding module and deep neural network module. In order to verify the performance of the proposed scheme, we implemented all the modules involved in the entire scheme. All modules are implemented in Python 2.7, and the performance is verified on a server with a GPU capacity of more than 10GB and a TensorFlow 1.4.0.

In the item graph construction module, we define two modes, static and dynamic, in fact, the two modes share the same storage structure. Because the directed weighted item graph is a large sparse graph, we use the form of adjacency list to store. In the adjacency list, each vertex has a singly-linked list. The node elements in the singly-linked list are related information of another vertex connected to the directed edge from the vertex. Each node contains the connected vertex, time weight and frequency weight. When in the static mode, the connected vertex and frequency weights of the vertices in the adjacency list will be activated; while in the dynamic mode, the connected vertices and time weights of the vertices in the adjacency list will be activated.

In the random walk module, we define two modes, static and dynamic. In the static mode, we set the length of random walk to 12 and the number of random walks per vertex to 20. By sampling 20 random walk sequences for each vertex in the item graph, we can obtain a set of item sequences with a maximum length of 12, which implies a high-order promixity relationship between items. In the dynamic mode, we set the length of the random walk to 12, and the start edge and start time are selected in an unbiased manner. Through the introduction of the time state, we can track the increase of the edge of the item graph according to the increase of the time stamp, thereby capturing the dynamic changes of the item graph. The random walk sequence sampled in a dynamic environment not only implies a high-order proximity relationship between items, but also allows the temporal relationship to be perfectly preserved according to the strict timing relationship, and can perform more temporal walks according to dynamic changes to meet System scalability requirements.

Static mode and dynamic mode share the same graph embedding module. The graph embedding module is mainly a shallow neural network model of Word2Vec. Specifically, the graph embedding module is a Skip-Gram model. In the Skip-Gram model, we set its embedding dimension to 180, the context window size to 20, the input is the item sequence, and the output is the item embedding vector. The essence of the entire graph embedding module is a mapping function, which can ensure that whether it is a static graph embedding vector or a dynamic graph embedding vector, if it has high-order proximity, then it appears to be close in the vector space.

\subsection{Experimental performance comparison}

We repeated all the comparison methods 8 times, and take the average of 5 results as our experimental results. Table~\ref{table3} shows the experimental results of various models on the Amazon electronic product data subset. 

In the static recommendation environment, all deep networks have greatly defeated the logistic regression model, which proves the effectiveness of deep learning to solve the recommendation problem. In the recommendation algorithm based on deep learning, PNN and DeepFM with special design structure are better than Wide\&Deep.However, we propose a static recommendation method based on graph embedding, S-DGEM, which performs best among all comparison methods. We owe it to the design graph embedding method in the static recommendation environment. The graph embedding design in the static recommendation environment can find high-order proximity between items through random walks with unequal probability, and the captured high-order proximity can be better weighted under the support of attention mechanism. Through this mechanism, S-DGEM can obtain an adaptive change representation of user interest, which greatly improves the expressive power of the model compared with other deep networks.

\begin{table}[!t]
	\renewcommand{\arraystretch}{1.3}
	\caption{Model comparison AUC and GAUC on the Amazon electronic product data subset. The static recommendation system calculates RelaImpr by comparing with BaseModel, and the dynamic recommendation system calculates RelaImpr by comparing with DeepWalk.}
	\label{table3}
	\centering
	\begin{tabular}{c|c|c|c|c}
		\hline
		\bfseries    & \bfseries AUC & \bfseries RelaImpr & \bfseries GAUC & \bfseries RelaImpr \\
		\hline
			BaseModel &	0.8635 &	0.0000\% &	0.8615 &	0.0000\% \\
		LR &	0.7759 &	-24.0990\% &	0.7741 &	-24.1770\% \\
		WideDeep &	0.8645 &	0.2751\% &	0.8622 &	0.1936\%
		\\
		PNN	 & 0.8658 &	0.6327\% &	0.8639 &	0.6639\% \\
		DeepFM &	0.8704 &	1.8982\% &	0.8684 &	1.9087\%
		\\
		S-DGEM &	0.8891	& 7.0426\% &	0.8869 &	7.0263\%
		\\
		DeepWalk &	0.8456 &	0.0000\% &	0.8319	& 0.0000\%
		\\
		LINE &	0.7366 &	-31.5394\% & 0.7202 &	-33.6547\%
		\\
		Node2Vec &	0.8607 &	4.3692\% &	0.8489 &	5.1220\%
		\\
		D-DGEM &	0.8571 &	3.3275\% &	0.8552 &	7.0202\% \\
		\hline
	\end{tabular}
\end{table}

In the dynamic recommendation environment, we note that temporal walks are time-biased and the random walk sampling has been biased towards the edges that appear later in time. Here, we see that D-DGEM always performs better than DeepWalk, Node2Vec and LINE. This means that when time information is ignored, important information is lost. What is confusing is that the performance of LINE does not show its advantages in this problem. The analysis reason is estimated that LINE does not take into account the higher order proximity and introduces a larger error. Compared with the static recommendation system, the performance of the AUC and GAUC of the dynamic recommendation system is reduced, but in view of introducing the time dimension that the static recommendation system does not have and achieving the scalability of the system, these performance losses are within an acceptable range. In other words, D-DGEM can make better use of the timing dependencies between items and improve the recommended performance in a dynamic environment. Description from the side, the inclusion of time dependencies in the graph is very important for learning an appropriate and meaningful network representation.

\begin{figure}[!t]
	\centering
	\includegraphics[scale=0.465]{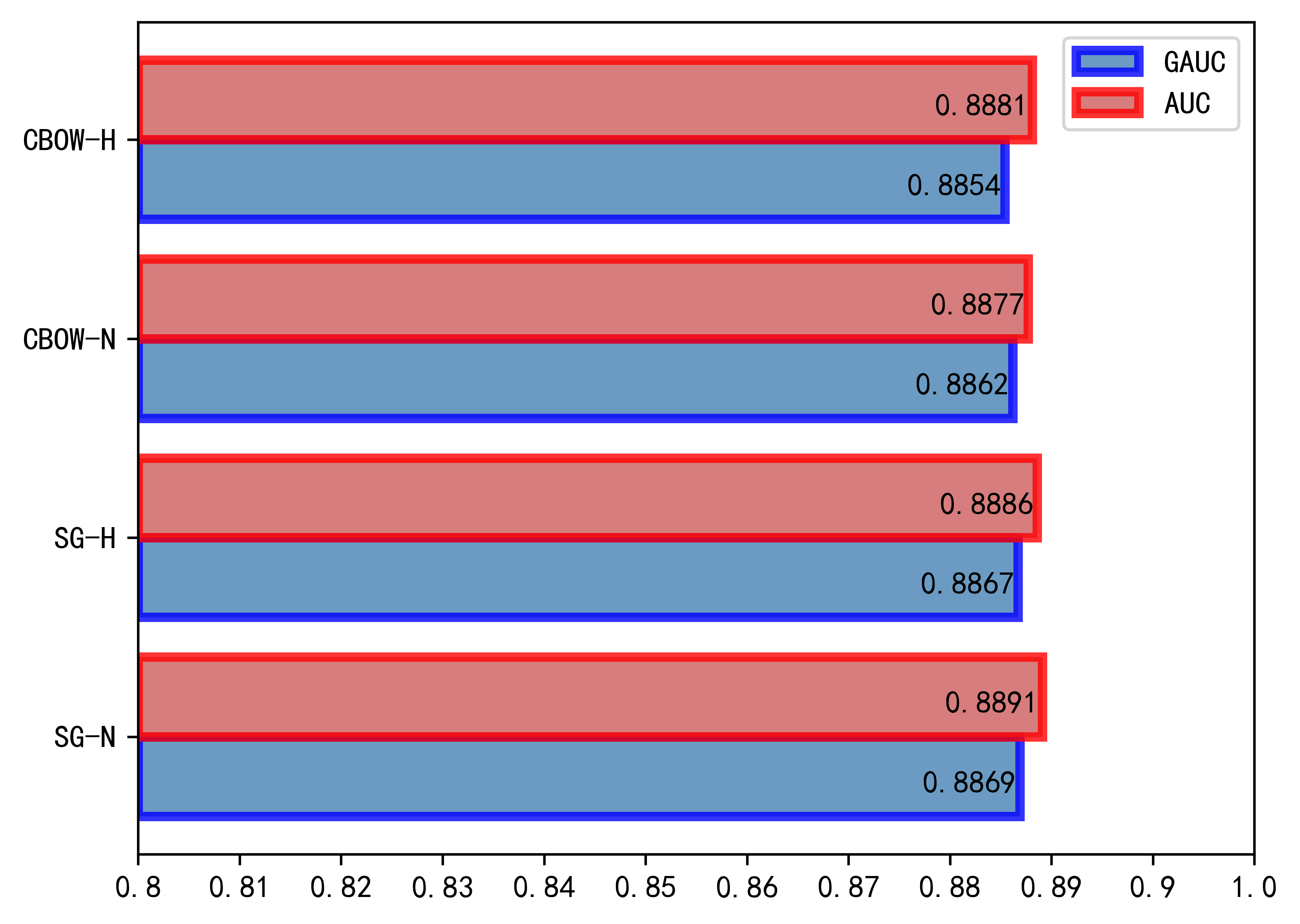}
	\caption{Effects of different graph embedding training methods on AUC and GAUC.}
	\label{fig7}
\end{figure}

The DGEM is a graph embedding model. When generating the graph embedding, the training model generally adopts the Word2Vec model. The Word2Vec model can be divided into Skip-Gram and CBOW, and the optimization methods are divided into negative sampling and Hierarchical softmax. Therefore, there are four methods for generating graph embedding, namely Skip-Gram with negative sampling(SG-N), Skip-Gram with Hierarchical softmax(SG-H), CBOW with negative sampling(CBOW-N) and CBOW with Hierarchical softmax(CBOW-H).

Fig.~\ref{fig7} show the AUC and GAUC for different graph embedding training methods on the Amazon electronic product data subset. We can find that the training method of graph embedding has little effect on the model. Skip-Gram with negative sampling method has the best effect on the training graph embedding model.

\begin{figure}[!t]
	\centering
	\includegraphics[scale=0.4]{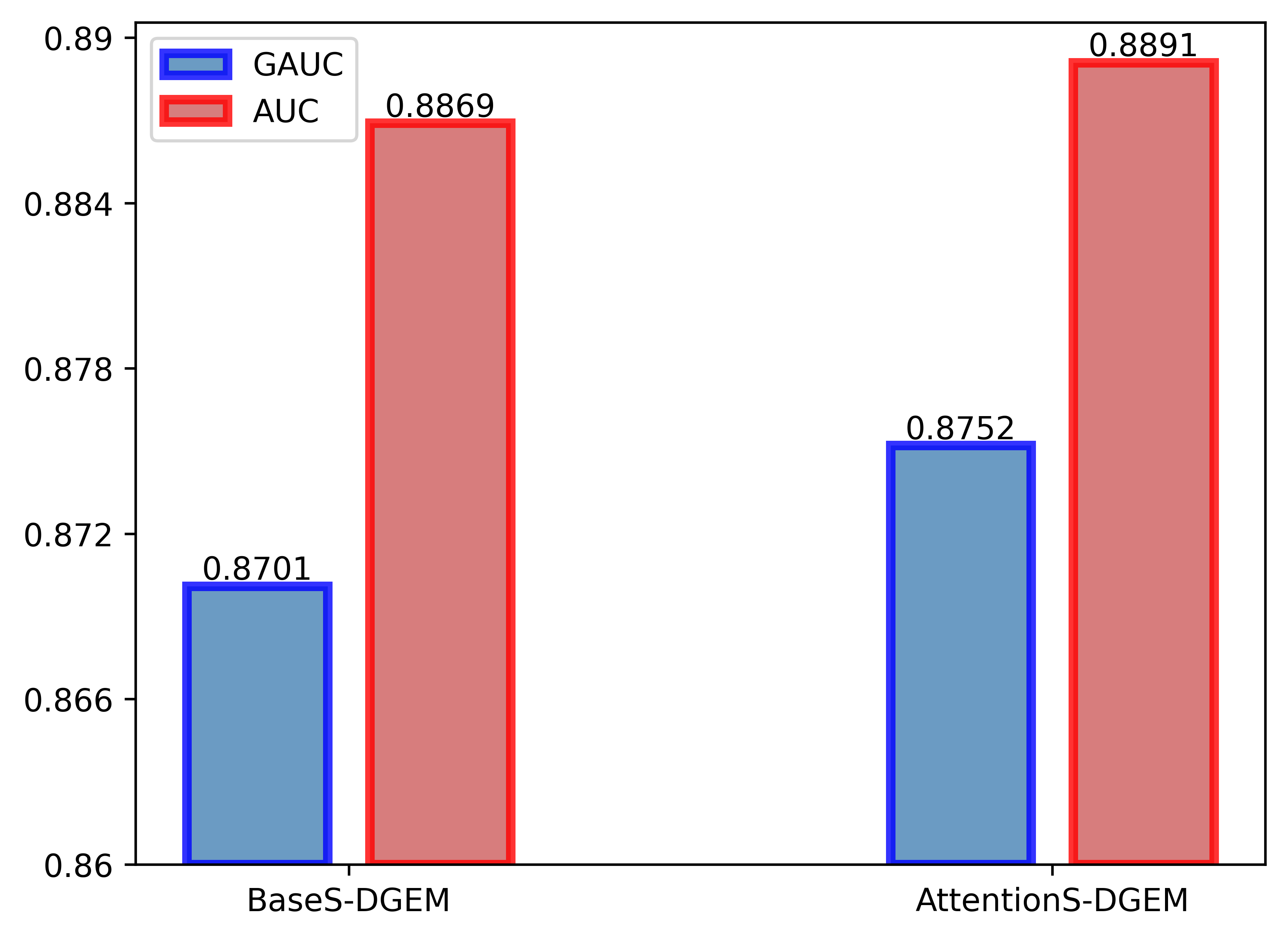}
	\caption{Effects of attention mechanism on AUC and GAUC.}
	\label{fig8}
\end{figure}

We also conducted experiments on the proposed attention mechanism. We conducted a comparative experiment between the pure static graph embedding method and the static graph embedding method with attention mechanism on the Amazon electronic product data subset. The experimental results are shown in Fig.~\ref{fig8}. By observing the experimental results, we can find that the static graph embedding method with the attention mechanism can obtain more competitive AUC and GAUC results. In summary, for our model, the static graph embedding recommendation method based on the attention mechanism can enhance the model's expressive ability and enhance the recommendation effect.

\begin{figure}[!t]
	\centering
	\includegraphics[scale=0.4]{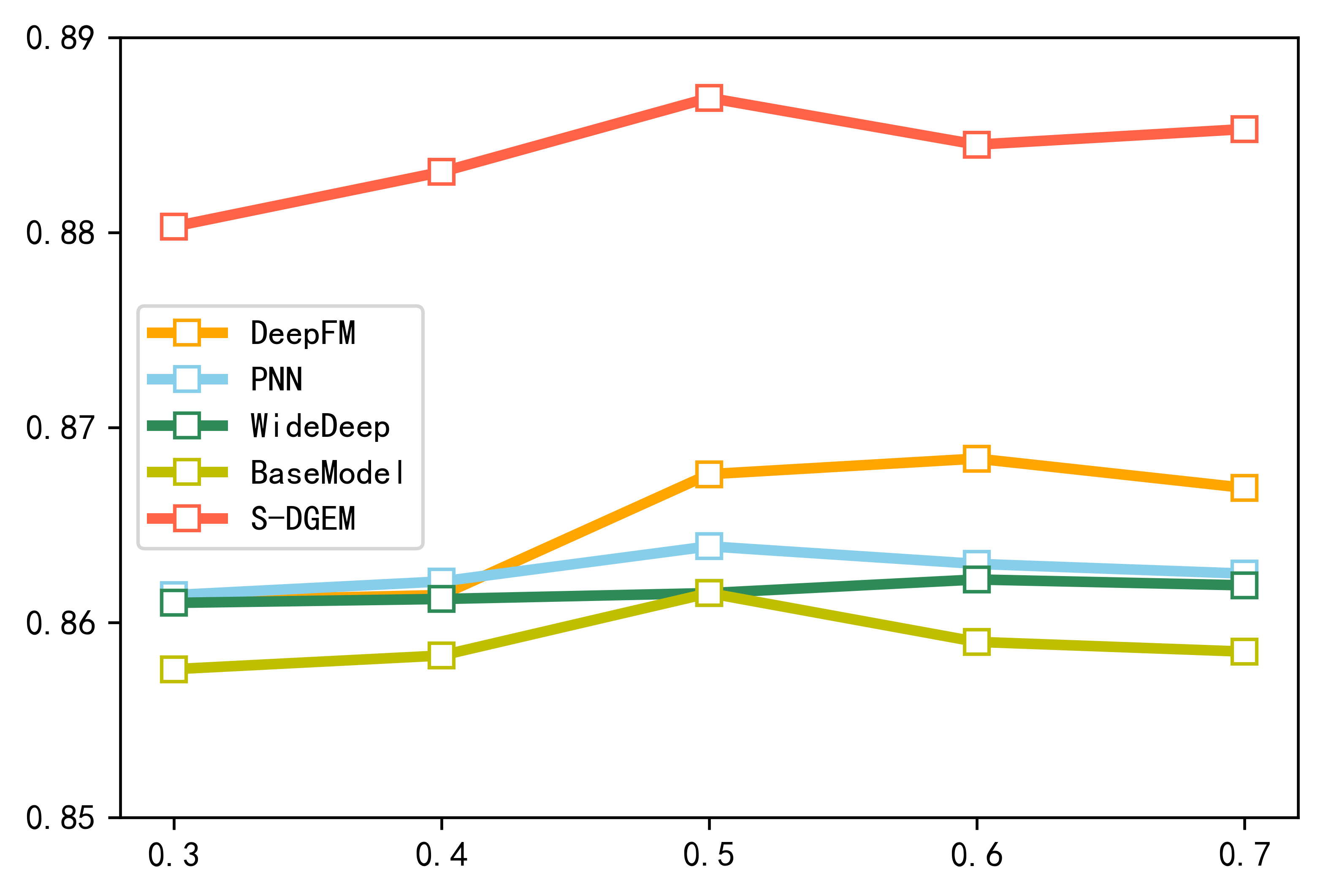}
	\caption{Effect of different Dropout probabilities on GAUC.}
	\label{fig9}
\end{figure}

\begin{figure}[!t]
	\centering
	\includegraphics[scale=0.4]{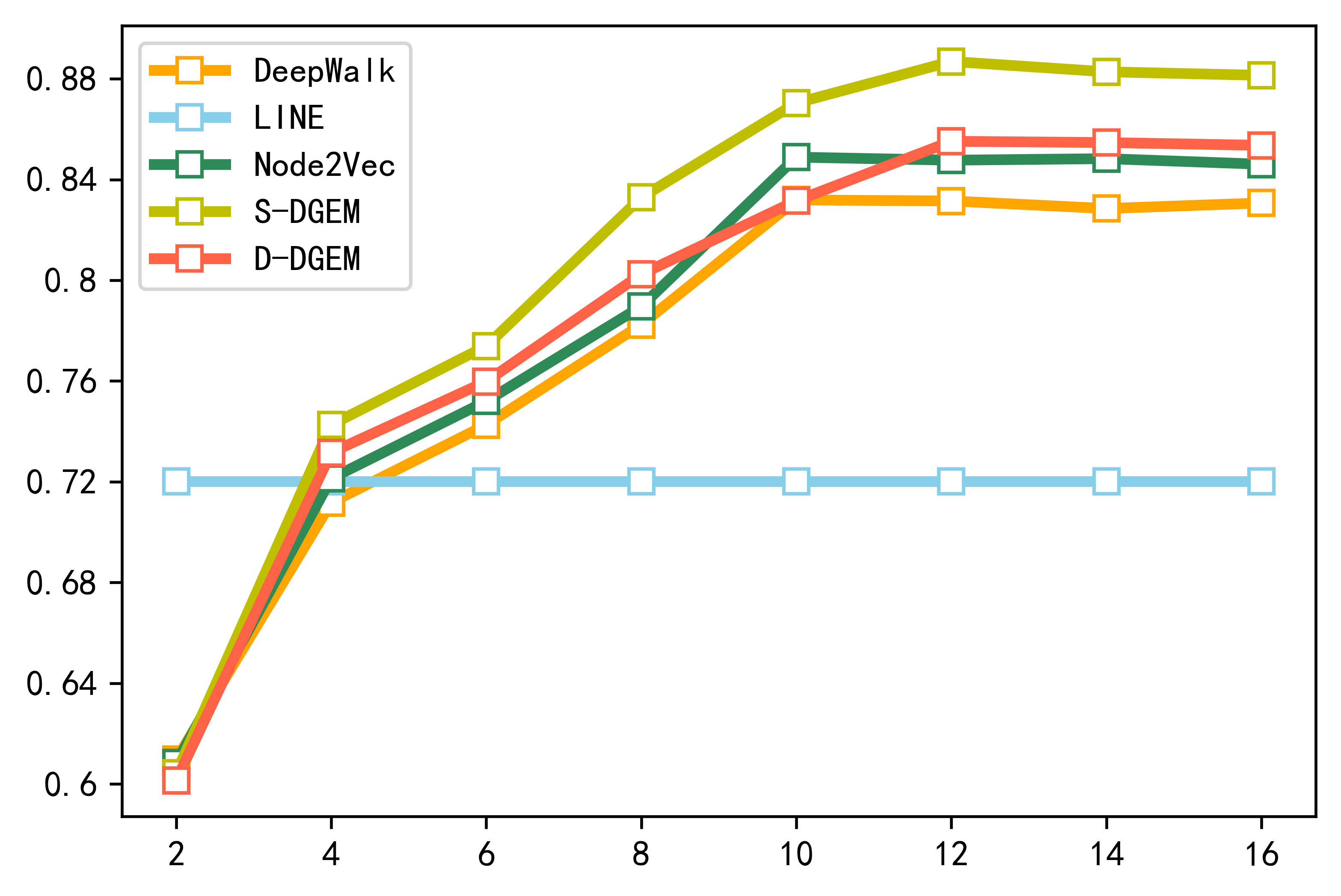}
	\caption{Effect of different length of random walk on GAUC.}
	\label{fig10}
\end{figure}

For the static graph embedding method, we mainly analyze the impact of different Dropout probabilities on GAUC; for the dynamic graph embedding recommendation method, we mainly analyze the impact of different length of random walk on GAUC. Fig.\ref{fig9} shows the experimental results of various static recommendation models using different Dropout probabilities on the Amazon electronic product data subset. Experimental results show that for all of static recommendation models, Dropout probability set at 0.5 or 0.6 can get the best recommendation effect. 

Fig.~\ref{fig10} shows the experimental results of various dynamic recommendation models using different length of random walk on the Amazon electronic product data subset. The experimental results show that when the length is 2, all the recommended models participating in the experiment are not effective, and with the increase of the length, except for the LINE, the recommendation effects of the other models are showing a steady growth trend. When the length is 10 or 12, the recommendation effect finally stabilizes. The reason why the experimental results of the LINE finally show a straight line is that LINE only considers the first-order proximity and second-order proximity between items.

\section{Conclusion}
\label{sec:conclusion}

In order to overcome the convergence problem caused by the simultaneous training of the embedding layer and the entire neural network and the problem that the current sequence embedding method is no longer applicable to the actual situation, our new technology solution DGEM is proposed. In a static recommendation environment, S-DGEM extracts the graph structure model by establishing a directed weighted item graph, uses random walks of unequal probability to capture the vertex attributes of the item graph, and generates item sequence data, and then generates the item graph embedding vector by the Word2Vec method , and finally feed the deep neural network for recommendation. The experimen- tal results show that S-DGEM can mine the high-order proximity between items and enhance the expression ability of the model.
In the dynamic recommendation environment, D-DGEM introduces the time state to track the update of the item graph, and improves the unequal probability random walk strategy in the static recommendation environment to capture the vertex attribute in the dynamic item graph, and adds auxiliary information to enhance the special solitary point expression. In this way, the timing dependence between items can be better utilized and the recommendation performance in a dynamic environment can be improved.

It is undeniable that the scale of the recommendation system to be processed is getting larger and larger, and the number of users and items participating in the recommendation system can easily reach 10 million. If the sparseness of the recommendation system is measured by the proportion of interactions among items in all possible interactions, the sparseness of the Amazon electronic product data set used in this article is only 0.0042\%. The problem of data sparsity is a big problem inherent in the recommendation system itself. This problem cannot be completely overcome in essence. In future work, we hope to be able to introduce more associations between items to better deal with data sparseness. In addition, we hope to design an adaptive algorithm in the future work. This algorithm can use better labels for tracking graph structure updates without generating cumulative errors and achieving true scalability.

% Can use something like this to put references on a page
% by themselves when using endfloat and the captionsoff option.
%\ifCLASSOPTIONcaptionsoff
%  \newpage
%\fi

\bibliographystyle{IEEEtran}  %要使用的格式,比如要投IEEE,就写IEEEtran
\bibliography{IEEEabrv,dgem_arxiv}{}%加载bi

% Generated by IEEEtran.bst, version: 1.13 (2008/09/30)
\begin{thebibliography}{10}
\providecommand{\url}[1]{#1}
\csname url@samestyle\endcsname
\providecommand{\newblock}{\relax}
\providecommand{\bibinfo}[2]{#2}
\providecommand{\BIBentrySTDinterwordspacing}{\spaceskip=0pt\relax}
\providecommand{\BIBentryALTinterwordstretchfactor}{4}
\providecommand{\BIBentryALTinterwordspacing}{\spaceskip=\fontdimen2\font plus
\BIBentryALTinterwordstretchfactor\fontdimen3\font minus
  \fontdimen4\font\relax}
\providecommand{\BIBforeignlanguage}[2]{{%
\expandafter\ifx\csname l@#1\endcsname\relax
\typeout{** WARNING: IEEEtran.bst: No hyphenation pattern has been}%
\typeout{** loaded for the language `#1'. Using the pattern for}%
\typeout{** the default language instead.}%
\else
\language=\csname l@#1\endcsname
\fi
#2}}
\providecommand{\BIBdecl}{\relax}
\BIBdecl

\bibitem{Word2Vec1}
T.~Mikolov, K.~Chen, G.~S. Corrado, and J.~Dean, ``Efficient estimation of word
  representations in vector space,'' in \emph{Proceedings of the 1st
  International Conference on Learning Representations(ICLR)}, 2013.

\bibitem{Word2Vec2}
T.~Mikolov, I.~Sutskever, K.~Chen, G.~Corrado, and J.~Dean, ``Distributed
  representations of words and phrases and their compositionality,'' in
  \emph{Proceedings of the 26th International Conference on Neural Information
  Processing Systems - Volume 2}, ser. NIPS’13, 2013, p. 3111–3119.

\bibitem{Word2Vec3}
X.~Rong, ``word2vec parameter learning explained,'' \emph{arXiv preprint
  arXiv:1411.2738}, 2014.

\bibitem{Item2Vec}
O.~{Barkan} and N.~{Koenigstein}, ``Item2vec: Neural item embedding for
  collaborative filtering,'' in \emph{2016 IEEE 26th International Workshop on
  Machine Learning for Signal Processing (MLSP)}, 2016, pp. 1--6.

\bibitem{Airbnb}
M.~Grbovic and H.~Cheng, ``Real-time personalization using embeddings for
  search ranking at airbnb,'' in \emph{Proceedings of the 24th ACM SIGKDD
  International Conference on Knowledge Discovery \& Data Mining}, ser. KDD
  ’18, 2018, p. 311–320.

\bibitem{Graph}
J.~Leskovec, J.~Kleinberg, and C.~Faloutsos, ``Graph evolution: Densification
  and shrinking diameters,'' \emph{ACM Trans. Knowl. Discov. Data}, vol.~1,
  no.~1, p. 2–es, Mar. 2007.

\bibitem{DeepWalk}
B.~Perozzi, R.~Al-Rfou, and S.~Skiena, ``Deepwalk: Online learning of social
  representations,'' in \emph{Proceedings of the 20th ACM SIGKDD International
  Conference on Knowledge Discovery and Data Mining}, 2014, p. 701–710.

\bibitem{LINE}
J.~Tang, M.~Qu, M.~Wang, M.~Zhang, J.~Yan, and Q.~Mei, ``Line: Large-scale
  information network embedding,'' in \emph{Proceedings of the 24th
  International Conference on World Wide Web}, ser. WWW ’15, 2015, p.
  1067–1077.

\bibitem{Node2Vec}
A.~Grover and J.~Leskovec, ``Node2vec: Scalable feature learning for
  networks,'' in \emph{Proceedings of the 22nd ACM SIGKDD International
  Conference on Knowledge Discovery and Data Mining}, ser. KDD ’16, 2016, p.
  855–864.

\bibitem{SDNE}
D.~Wang, P.~Cui, and W.~Zhu, ``Structural deep network embedding,'' in
  \emph{Proceedings of the 22nd ACM SIGKDD International Conference on
  Knowledge Discovery and Data Mining}, ser. KDD ’16, 2016, p. 1225–1234.

\bibitem{IO1}
D.~Rosenberg, ``Early modern information overload,'' \emph{Journal of the
  History of Ideas}, vol.~64, pp. 1--9, 01 2003.

\bibitem{IO2}
A.~Borchers, J.~Herlocker, J.~Konstan, and J.~Riedl, ``Ganging up on
  information overload,'' \emph{Computer}, vol.~31, no.~4, p. 106–108, Apr.
  1998.

\bibitem{survey}
K.~{Wei}, J.~{Huang}, and S.~{Fu}, ``A survey of e-commerce recommender
  systems,'' in \emph{2007 International Conference on Service Systems and
  Service Management}, 2007, pp. 1--5.

\bibitem{LE}
M.~Belkin and P.~Niyogi, ``Laplacian eigenmaps and spectral techniques for
  embedding and clustering,'' in \emph{Proceedings of the 14th International
  Conference on Neural Information Processing Systems: Natural and Synthetic},
  ser. NIPS’01, 2001, p. 585–591.

\bibitem{LLE}
S.~T. Roweis and L.~K. Saul, ``Nonlinear dimensionality reduction by locally
  linear embedding,'' \emph{Science}, vol. 290, no. 5500, pp. 2323--2326, 2000.

\bibitem{GF}
A.~Ahmed, N.~Shervashidze, S.~Narayanamurthy, V.~Josifovski, and A.~J. Smola,
  ``Distributed large-scale natural graph factorization,'' in \emph{Proceedings
  of the 22nd International Conference on World Wide Web}, ser. WWW ’13,
  2013, p. 37–48.

\bibitem{GraRep}
\BIBentryALTinterwordspacing
S.~Cao, W.~Lu, and Q.~Xu, ``Grarep: Learning graph representations with global
  structural information,'' in \emph{Proceedings of the 24th ACM International
  on Conference on Information and Knowledge Management}, ser. CIKM ’15,
  2015, p. 891–900. [Online]. Available:
  \url{https://doi.org/10.1145/2806416.2806512}
\BIBentrySTDinterwordspacing

\bibitem{HOPE}
M.~Ou, P.~Cui, J.~Pei, Z.~Zhang, and W.~Zhu, ``Asymmetric transitivity
  preserving graph embedding,'' in \emph{Proceedings of the 22nd ACM SIGKDD
  International Conference on Knowledge Discovery and Data Mining}, ser. KDD
  ’16, 2016, p. 1105–1114.

\bibitem{centrality}
M.~J. Newman, ``A measure of betweenness centrality based on random walks,''
  \emph{Social Networks}, vol.~27, no.~1, pp. 39 -- 54, 2005.

\bibitem{similarity}
F.~{Fouss}, A.~{Pirotte}, J.~{Renders}, and M.~{Saerens}, ``Random-walk
  computation of similarities between nodes of a graph with application to
  collaborative recommendation,'' \emph{IEEE Transactions on Knowledge and Data
  Engineering}, vol.~19, no.~3, pp. 355--369, 2007.

\bibitem{HARP}
H.~Chen, B.~Perozzi, Y.~Hu, and S.~Skiena, ``Harp: Hierarchical representation
  learning for networks,'' 06 2017.

\bibitem{Walklets}
B.~Perozzi, V.~Kulkarni, and S.~Skiena, ``Walklets: Multiscale graph embeddings
  for interpretable network classification,'' \emph{CoRR}, vol. abs/1605.02115,
  2016.

\bibitem{GenVector}
Z.~Yang, J.~Tang, and W.~Cohen, ``Multi-modal bayesian embeddings for learning
  social knowledge graphs,'' in \emph{Proceedings of the Twenty-Fifth
  International Joint Conference on Artificial Intelligence}, ser. IJCAI’16,
  2016, p. 2287–2293.

\bibitem{DDRW}
J.~Li, J.~Zhu, and B.~Zhang, ``Discriminative deep random walk for network
  classification,'' in \emph{Proceedings of the 54th Annual Meeting of the
  Association for Computational Linguistics (Volume 1: Long Papers)}.\hskip 1em
  plus 0.5em minus 0.4em\relax Berlin, Germany: Association for Computational
  Linguistics, Aug. 2016, pp. 1004--1013.

\bibitem{TriDNR}
S.~Pan, J.~Wu, X.~Zhu, C.~Zhang, and Y.~Wang, ``Tri-party deep network
  representation,'' in \emph{Proceedings of the Twenty-Fifth International
  Joint Conference on Artificial Intelligence}, ser. IJCAI’16, 2016, p.
  1895–1901.

\bibitem{other}
Z.~Yang, W.~W. Cohen, and R.~Salakhutdinov, ``Revisiting semi-supervised
  learning with graph embeddings,'' in \emph{Proceedings of the 33rd
  International Conference on International Conference on Machine Learning -
  Volume 48}, ser. ICML’16, 2016, p. 40–48.

\bibitem{LR}
H.~B. McMahan, G.~Holt, D.~Sculley, M.~Young, D.~Ebner, J.~Grady, L.~Nie,
  T.~Phillips, E.~Davydov, D.~Golovin, S.~Chikkerur, D.~Liu, M.~Wattenberg,
  A.~M. Hrafnkelsson, T.~Boulos, and J.~Kubica, ``Ad click prediction: a view
  from the trenches,'' in \emph{Proceedings of the 19th ACM SIGKDD
  International Conference on Knowledge Discovery and Data Mining (KDD)}, 2013.

\bibitem{WideDeep}
H.-T. Cheng, L.~Koc, J.~Harmsen, T.~Shaked, T.~Chandra, H.~Aradhye,
  G.~Anderson, G.~Corrado, W.~Chai, M.~Ispir, R.~Anil, Z.~Haque, L.~Hong,
  V.~Jain, X.~Liu, and H.~Shah, ``Wide \& deep learning for recommender
  systems,'' in \emph{Proceedings of the 1st Workshop on Deep Learning for
  Recommender Systems}, ser. DLRS 2016, 2016, p. 7–10.

\bibitem{PNN}
Y.~Qu, B.~Fang, W.~Zhang, R.~Tang, M.~Niu, H.~Guo, Y.~Yu, and X.~He,
  ``Product-based neural networks for user response prediction over multi-field
  categorical data,'' \emph{ACM Trans. Inf. Syst.}, vol.~37, no.~1, Oct. 2018.

\bibitem{DeepFM}
H.~Guo, R.~Tang, Y.~Ye, Z.~Li, and X.~He, ``Deepfm: A factorization-machine
  based neural network for ctr prediction,'' in \emph{Proceedings of the 26th
  International Joint Conference on Artificial Intelligence}, ser. IJCAI’17,
  2017, p. 1725–1731.

\end{thebibliography}

% biography section
% 
% If you have an EPS/PDF photo (graphicx package needed) extra braces are
% needed around the contents of the optional argument to biography to prevent
% the LaTeX parser from getting confused when it sees the complicated
% \includegraphics command within an optional argument. (You could create
% your own custom macro containing the \includegraphics command to make things
% simpler here.)
%\begin{IEEEbiography}
% or if you just want to reserve a space for a photo:

% You can push biographies down or up by placing
% a \vfill before or after them. The appropriate
% use of \vfill depends on what kind of text is
% on the last page and whether or not the columns
% are being equalized.

%\vfill

% Can be used to pull up biographies so that the bottom of the last one
% is flush with the other column.
%\enlargethispage{-5in}

% that's all folks
\end{document}